\documentclass[12pt]{article}
\newtheorem{lemma}{Lemma}
\newtheorem{proposition}{Proposition}
\newtheorem{theorem}{Theorem}
\newtheorem{corollary}{Corollary}
\newtheorem{definition}{Definition}
\def\tr{\mathop{\rm Tr}\nolimits}

\def\ci{\mathop{\textrm{i}}\nolimits}

\catcode`\@=11 \long\def\@makefntext#1{ \protect\noindent \hbox to
3.2pt {\hskip-.9pt
$^{{\ninerm\@thefnmark}}$\hfil}#1\hfill}        

 \def\@makefnmark{\hbox to 0pt{$^{\@thefnmark}$\hss}}  

\def\ps@myheadings{\let\@mkboth\@gobbletwo
\def\@oddhead{\hbox{}
\rightmark\hfil\ninerm\thepage}
\def\@oddfoot{}\def\@evenhead{\ninerm\thepage\hfil
\leftmark\hbox{}}\def\@evenfoot{}
\def\sectionmark##1{}\def\subsectionmark##1{}}

\renewenvironment{thebibliography}[1]
    {\begin{list}{$^{\arabic{enumi}}$}
    {\usecounter{enumi}\setlength{\parsep}{0pt}
\setlength{\leftmargin 1.25cm}{\rightmargin 0pt}
     \setlength{\itemsep}{0pt} \settowidth
    {\labelwidth}{#1.}\sloppy}}{\end{list}}

\newcounter{itemlistc}
\newcounter{romanlistc}
\newcounter{alphlistc}
\newcounter{arabiclistc}

\def\@citex[#1]#2{\if@filesw\immediate\write\@auxout
    {\string\citation{#2}}\fi
\def\@citea{}\@cite{\@for\@citeb:=#2\do
    {\@citea\def\@citea{,}\@ifundefined
    {b@\@citeb}{{\bf ?}\@warning
    {Citation `\@citeb' on page \thepage \space undefined}}
    {\csname b@\@citeb\endcsname}}}{#1}}

\newif\if@cghi
\def\cite{\@cghitrue\@ifnextchar [{\@tempswatrue
    \@citex}{\@tempswafalse\@citex[]}}
\def\citelow{\@cghifalse\@ifnextchar [{\@tempswatrue
    \@citex}{\@tempswafalse\@citex[]}}
\def\@cite#1#2{{$\null^{#1}$\if@tempswa\typeout
    {IJCGA warning: optional citation argument
    ignored: `#2'} \fi}}

\def\fnt#1#2{\footnotetext{\kern-.3em
    {$^{\mbox{\sevenrm #1}}$}{#2}}}

 1 
scaled\magstep 1  1
 
scaled\magstephalf 
  
 \font\ninerm=cmr9

\begin{document}

\title{On the geometry of Killing and conformal tensors}
\author{Bartolom\'e Coll$^1$,\
Joan Josep Ferrando$^2$\ and Juan Antonio S\'aez$^3$}
\date{\empty}

\maketitle \vspace*{-0.5cm}

\begin{abstract}
The second order Killing and conformal tensors are analyzed in terms
of their spectral decomposition, and some properties of the
eigenvalues and the eigenspaces are shown. When the tensor is of
type I with only two different eigenvalues, the condition to be a
Killing or a conformal tensor is characterized in terms of its
underlying almost-product structure. A canonical expression for the
metrics admitting these kinds of symmetries is also presented. The
space-time cases 1+3 and 2+2 are analyzed in more detail. Starting
from this approach to Killing and conformal tensors a geometric
interpretation of some results on quadratic first integrals of the
geodesic equation in vacuum Petrov-Bel type D solutions is offered.
A generalization of these results to a wider family of type D
space-times is also obtained.
\end{abstract}

\vspace*{2mm}
\begin{center}
\end{center}

\vspace*{8mm} \noindent $^1$\ Syst\`emes de r\'ef\'erence
relativistes, SYRTE-CNRS, Obsevatoire de Paris, 75014 Paris, France. E-mail: {\tt bartolome.coll@obspm.fr} \\
$^2$ Departament d'Astronomia i Astrof\'{\i}sica, Universitat de
Val\`encia, E-46100 Burjassot, Val\`encia, Spain.
E-mail: {\tt joan.ferrando@uv.es}\\
$^3$ Departament de Matem\`atiques per a l'Economia i l'Empresa,
Universitat de Val\`encia, E-46071 Val\`encia, Spain. E-mail: {\tt
juan.a.saez@uv.es}
\newpage

\section{\small INTRODUCTION}

Killing tensors are associated with first integrals to the geodesic
equation. In the second order case, they define quadratic first
integrals and they play a central role in the theory of separability
of the Hamilton-Jacobi equation. The relationship between
separability and Killing tensors was shown by Eisenhart\cite{eis}
and abundant literature exists regarding this property (for example,
see Ref. 2 and references therein).

Within the relativistic framework the study of Killing tensors grew
when Walker and Penrose\cite{wape} showed how the existence of a
Killing tensor explains the Carter results\cite{car} on the
integrability by variable separation of the geodesic equation in the
Kerr solution. Since then a lot of studies have been devoted to
determining and classifying the space-times admitting Killing
tensors and also to obtaining the Killing tensors of a given metric.
A summary of known results on this subject can be found in Ref. 5.

The problem of finding the metrics admitting a quadratic integral of
the geodesic equation was established by Eisenhart.\cite{eis} He
wrote the {\it intrinsic Killing tensor equations}, i.e., the
Killing equations in terms of the eigenvectors $e_i$ and the
eigenvalues $\rho_i$ of a Killing tensor, and he pointed out that
(see Ref. 1, pag. 129): "the problem of finding all $V_n$ admitting
a quadratic integral consists in finding a tensor $g$ and an
orthogonal ennuple $e_i$ that satisfy the conditions obtained by the
elimination of the $\rho$'s from the intrinsic Killing tensor
equations. The general solution has not been obtained, but we shall
consider two particular solutions of the problem". Later, he
considered the trivial case when all the $\rho$'s are equal, and the
case with different eigenvalues and normal principal congruences, a
case which led to the St\"{a}ckel form of the metric.\cite{sta,eis}

The general solution to the problem set by Eisenhart is far from
being solved, although a number of results are known for some
classes of Einstein-Maxwell solutions or algebraically special
space-times, as well as those for flat metrics.\cite{kramer}
Nevertheless, the usual way in which this subject is tackled differs
from the Eisenhart conception. Indeed, the common approach consists
of studying the integrability conditions of the Killing tensor
equations, whereas the Eisenhart method involves the following: (i)
to write the intrinsic Killing tensor equations, (ii) to determine
the equivalent equations involving exclusively the eigenspaces and
the metric tensor (the eigenvalues having been removed), and (iii)
to study the integrability conditions of the aforementioned
equations. Both procedures, the usual one and Eisenhart's, may be
suitable depending on the different situations. In this work we
adopt the Eisenhart approach and we will show how useful it is by
considering the case of Killing tensors with two complementary
eigenspaces.

The conformal extension of the Killing tensor equation determines
the conformal tensors which define first integrals to the null
geodesic equation. Here we also analyze the Eisenhart problem for
the class of conformal tensors with two complementary eigenspaces.

In the problem of finding the Riemannian spaces admitting a Killing
or a conformal tensor two different aspects can be considered. On
one hand, we can look for a general canonical expression for the
metric tensors with these kinds of first integrals. In this case, we
must also obtain the expression of the Killing or conformal tensors
in terms of the elements appearing in this canonical form. This
approach may be useful in working in spaces with these symmetries,
the adapted coordinates allowing calculations to be simplified and
throwing light on the geometric interpretation of the expressions we
can find.

On the other hand, we can give explicit and intrinsic conditions
that characterize the metric tensors, and then we must offer the
expression of the Killing or conformal tensors in terms of metric
concomitants (namely, the Riemann tensor and its covariant
derivatives). This approach is helpful in analyzing when a metric,
which is known in an arbitrary coordinate system, has these kinds of
symmetries. Moreover, we can obtain these tensorial symmetries
without solving the Killing or conformal equations.

In this work we analyze both viewpoints. Regarding the first one, we
can quote several results previously obtained in the relativistic
framework. Thus, canonical forms for the four dimensional space-time
metrics admitting a Killing or a conformal tensor of type $2+2$ have
been proposed in literature.\cite{h-m-1,h-m-2} In this case the
Killing or conformal tensor admits two complementary eigenplanes.
Here we generalize these results by considering a general $p+q$
tensor (with two complementary eigenspaces of dimensions $p$ and
$q$, respectively) in a generic Riemannian space with arbitrary
signature and dimension.

The second approach, the intrinsic characterization of the metrics
admitting Killing and conformal tensors, has also been partially
considered in relativity. Thus, it is known that every Petrov-Bel
type D vacuum solution admits a conformal tensor of type $2+2$ which
may be obtained form the Weyl tensor.\cite{kramer} Here we extend
this result by characterizing all the Petrov-Bel type D metrics with
conformal tensors. Moreover we also identify the type D solutions
admitting a Killing tensor, thus generalizing some results that are
known for the vacuum case.\cite{kramer}

It is worth remarking that the Eisenhart approach used here allows
the intrinsic and explicit labeling of the metrics to be obtained
easily. Indeed, in this approach we give conditions for the
underlying $2+2$ structure of the Killing or conformal tensors.
Moreover, for the Petrov-Bel type D metrics, this is the principal
structure one of the Weyl tensor, and it is explicitly known in
terms of the metric tensor.\cite{fms} The reason why it is of
interest to obtain an explicit and intrinsic characterization of a
space-time metric has been pointed out elsewhere\cite{fsS} and the
method used here has been useful in labeling the
Schwarszchild\cite{fsS} and Reissner-Nordstr\"{o}m\cite{fsD}
solutions, the static Petrov type I space-times\cite{fms} and the
Petrov type I space-times admitting isotropic radiation.\cite{fsI}

Here we show that the eigenspaces of a Killing or a conformal tensor
are umbilical planes. Moreover they are totally geodesic for a
conformal metric. This geometric interpretation could be useful in
clarifying the role played by the Killing tensor in the separability
theory.

The paper is organized as follows. Some notation, definitions and
properties related to regular Riemannian p-planes are introduced in
section 2. In section 3 we study some properties of the eigenvectors
and eigenvalues of a Killing or a conformal tensor. The type I case
(when the tensor admits an orthonormal basis of eigenvectors) is
analyzed in detail in section 4 and we write the Eisenhart intrinsic
Killing tensor equations in a form that is more useful to our
purposes. In section 5 we use this new form for the Killing tensor
equations to analyze the Eisenhart problem when the Killing or the
conformal tensor has two complementary eigenspaces. A canonical form
for the metrics admitting these kinds of first integrals is
presented in section 6. In section 7 we study the 1+(n-1) case and
outline when these Killing or conformal tensors are not reducible.
In the last two sections some results concerning the usual four
dimensional space-time are obtained. The 2+2 space-time structures
associated to a Killing or conformal tensor are analyzed in detail
in section 8. Finally, section 9 is devoted to obtaining an
intrinsic and explicit characterization of the Petrov-Bel type D
metrics admitting Killing or conformal tensors attached to its
principal structure, and we also present an algorithm to obtain
these quadratic first integrals in a given type D space-time.

\section{\small SOME NOTATION AND USEFUL CONCEPTS}
\label{sec-notation}

On an $n$--dimensional Riemannian manifold $(M,g)$ we shall refer to
a (regular) $p$-dimensional distribution $V$ as a $p$--plane. Let
$v$ be the projector on $V$ and $h=g-v$ the projector on the plane
orthogonal to $V$. The generalized second fundamental form of $V$ is
defined as the (2,1)-tensor $Q_v$ given by
\begin{equation} \label{sff1}
Q_v (x,y) = h (\nabla_{v(x)} v(y))
\end{equation}
for every pair of vector fields $x$, $y$. We can consider the
decomposition of $Q_v$ into its antisymmetric part $A_v$ and its
symmetric part $S_v \equiv S_v^T + {1 \over p}v \otimes \tr S_v$,
where $S_v^T$ is a traceless tensor:
\begin{equation}  \label{Q2}
Q_v = A_v + \frac{1}{p} v \otimes \tr S_v + S_v^T
\end{equation}
The plane $V$ is a foliation if, and only if, $A_v =0$. In this case
$Q_v = S_v$ and it coincides with the second fundamental form of the
integral manifolds of the foliation $V$.\cite{rei} Moreover $V$ is
minimal, umbilical or geodesic if, and only if, $\tr S_v=0$, $S_v^T
=0$ or $S_v  =0$, respectively. Then one can generalize these
geometric concepts for plane fields which are not necessarily a
foliation:

\begin{definition}
A plane field $V$ is said to be geodesic, umbilical or minimal if
the symmetric part $S_v$ of its (generalized) second fundamental
form $Q_v$ satisfies $S_v =0$, $S_v^T =0$ or $\tr S_v =0$,
respectively.
\end{definition}
From these definitions, and defining $\{x,y\} = \nabla_x y +
\nabla_y x$, a lemma easily follows:

\begin{lemma} \label{lemma-umbilical}
A plane field $V$ is umbilical for the metric $g$ if, and only if, a
vector field $\textit{\textbf a}$ exists such that $\, h(\{ x, y \})
= g(x,y) \, \textit{\textbf a} \, $ for every $x, y \in V$, $h$
being the projector on the plane orthogonal to $V$.
\end{lemma}

On a $n$--dimensional Riemannian manifold $(M,g)$ an almost-product
structure is defined by a p-plane field $V$ and its orthogonal
complement $H$. The almost-product structures can be classified
taking into account the invariant decomposition of the covariant
derivative of the {\it structure tensor} $\, \Pi =v-h \,$. Likewise,
they can be classified according to the foliation, minimal,
umbilical or geodesic character of each plane.\cite{nav,olga}  We
will say that a structure $(V,H)$ is integrable when both planes are
foliations and we will say that it is minimal, umbilical or geodesic
if both of the planes are so.

In an oriented four dimensional space-time $(V_4,g)$ of signature
$(-+++)$ a more accurate classification for the almost-product
structures follows taking into account the causal character of the
planes.\cite{cf} Elsewhere\cite{fsD} we have classified the
Petrov-Bel type D space-times in accordance with the class of the
$2+2$ principal structure of the Weyl tensor.

\section{\small SECOND ORDER KILLING AND CONFORMAL TENSORS}
\label{sec-KCT}

The quadratic first integrals of the geodesic equation are
associated with second rank {\em Killing tensors}.\cite{eis}
Indeed, if $K$ is a solution to the generalized {\em Killing
equation}
\begin{equation} \label{Kd}
[K, g] = 0   \qquad  \qquad  ([K,g]_{abc} = \nabla_{(a} K_{bc)}) \
,
\end{equation}
then the scalar $K(v,v)$ is constant along an affine parameterized
geodesic with tangent vector $v$.

It is known\cite{kramer} that if $K$ is a Killing tensor, its
traceless part $\displaystyle P= K - \frac{1}{n} \tr K g$ is a {\em
conformal tensor}, i.e. it satisfies the {\em conformal equation}:
\begin{equation} \label{Cd}
[P,g] = {\cal S}\{g \otimes t\}
\end{equation}
where $t$ is, up to a factor, the divergence of $P$,
$t=\frac{2}{n+2}\nabla \cdot{P}$, and ${\cal S}\{B\}$ denotes the
total symmetrization of a tensor $B$. Then, the scalar $P(v,v)$ is
constant along an affinely parameterized null geodesic with
tangent vector $v$. Moreover, Killing equation (\ref{Kd}) implies:
\begin{equation}
 2 n \nabla \cdot P + (n+2) {\rm d} \tr K = 0     \label{Kd2}
\end{equation}
Then, we have the following

\begin{lemma} \label{lem-K-C}
If $K$ is a second rang Killing tensor (solution to {\rm
(\ref{Kd})}) then its traceless part $\displaystyle P= K -
\frac{1}{n} \tr K g$ is a conformal tensor (solutions to {\rm
(\ref{Cd})}) and it satisfies
\begin{equation}
{\rm d} \nabla \cdot P = 0                \label{Kd3}
\end{equation}
Conversely if a traceless conformal tensor $P$ satisfies {\rm
(\ref{Kd3})}, a scalar $\pi$ exists such that ${\rm d} \pi = \nabla
\cdot P$. Then, $K= P - \frac{2}{n+2} \pi g$ is a Killing tensor.
\end{lemma}

In this work we analyze some properties of the eigenvalues and
eigenspaces of Killing and conformal tensors and we present some of
their properties. We proceed by studying both classes of tensors
simultaneously and we will comment on the differences when they
exist. So, if we consider a second rang tensor $T$ solution to
(\ref{Cd}) the consequences on its eigenspaces and eigenvalues apply
to both, Killing and conformal tensors. We particularize the
conformal case by taking $T$ as a traceless tensor. If we add
condition (\ref{Kd3}), then $T$ is the traceless part of a Killing
tensor. But we can also recover the Killing tensor case by taking
the vector $t$ to be zero. It is worth pointing out that if $P$ is a
traceless conformal tensor, then $P+ \Phi g$ is a conformal tensor,
and both define the same first integrals of the null geodesic
equation. Nevertheless, here we will always work with the traceless
representative.

We denote $E_{\rho}$ the eigenspace of $T$ corresponding to the
eigenvalue $\rho$. Then, if  $x,y \in E_{\rho}$, a straightforward
calculation leads to:
\begin{equation}
[T,g](x,y,\cdot) = x(\rho) y + y(\rho) x + g(x,y) {\rm d} \rho -
(T- \rho g) \{x,y\}
\end{equation}
On the other hand,
\begin{equation}
{\cal S}\{g \otimes t\}(x,y,\cdot) = g(x,y) t + g(t,x)y + g(t,y)x
\end{equation}
So, for two eigenvectors $x,y \in E_{\rho}$, the conformal
condition (\ref{Cd}) implies:
\begin{equation}  \label{eigen1}
 (T- \rho g) \{x,y\} = g(x,y) s + g(s,x)y + g(s,y)x \, ,
 \qquad s \equiv {\rm d} \rho  - t
\end{equation}

On the other hand, if we consider three eigenvectors $x,y,z$
corresponding to three different eigenvalues, a similar calculation
leads to:
\begin{equation}  \label{eigen2}
T(x, \{y,z\}) + T(z, \{x,y\}) + T(y, \{z,x\}) = 0
\end{equation}
Thus, we can state the following:

\begin{lemma}  \label{lem-eigen}
Let $T$ be a Killing (respectively, conformal) tensor. Then:

(i) If $x,y \in E_{\rho}$ are eigenvectors associated with the
eigenvalue $\rho$, equation {\rm (\ref{eigen1})} holds, where the
vector $t$ is zero (respectively, $t=\frac{2}{n+2}\nabla \cdot{T}$).

(ii) If $x,y,z$ are eigenvectors corresponding to three different
eigenvalues, equation {\rm (\ref{eigen2})} holds.
\end{lemma}

A consequence of lemma \ref{lem-eigen} follows by taking $x=y$ in
equation (\ref{eigen1}). Indeed, if one makes a new product with $x$
one obtains:
\begin{equation}  \label{eigen3}
x^2 g({\rm d} \rho  - t, x) = 0
\end{equation}
and so, if $x,y$ are non null vectors, equation (\ref{eigen1})
becomes:
\begin{equation}  \label{eigen1b}
 (T- \rho g) \{x,y\} = g(x,y) ({\rm d} \rho  - t)
\end{equation}

If $E_{\rho}$ is a regular eigenspace of $T$, then a basis of
$E_{\rho}$ formed with non null eigenvectors exists and,
consequently, (\ref{eigen1b}) holds even for the null eigenvectors.
Moreover, taking into account (\ref{eigen3}) we have:

\begin{lemma}  \label{lem-regular}
Let $E_{\rho}$ be a regular eigenspace of a Killing (respectively,
conformal) tensor $T$. Then {\rm (\ref{eigen1b})} with $t=0$
(respectively,
$t=\frac{2}{n+2}\nabla \cdot{T}$) holds for every $x,y \in E_{\rho}$.\\[1mm]
Moreover $\, {\rm d} \rho  \in E_{\rho}^{\perp} \, $ (respectively,
$\,  2 \nabla \cdot{T} - (n+2){\rm d} \rho \in E_{\rho}^{\perp} \,
$).
\end{lemma}

\section{\small EIGENVALUES AND EIGENVECTORS OF SECOND ORDER
KILLING AND CONFORMAL TENSORS OF TYPE I}  \label{sec-KCT-I}

Let us now go to type I Killing and conformal tensors, that is those
admitting an orthonormal basis of eigenvectors. In this case every
eigenspace is regular and then the Killing (or conformal) equation
implies (\ref{eigen2}) and (\ref{eigen1b}). Moreover a basis of
eigenvectors exists and, consequently, these restrictions are also
sufficient conditions for T to be a Killing (or conformal) tensor.
Thus, we have:
\begin{proposition} \label{caracter1}
Let $T$ be a symmetric 2-tensor of type I and let $E_{i}$ be the
eigenspaces corresponding to the eigenvalues $\rho_i$. Then, $T$ is
a Killing (respectively, conformal) tensor if, and only if:
\begin{description}
\item (i)  $(T- \rho_i g) \{x,y\} = g(x,y) ({\rm d} \rho_i  - t)$,
for every $x, y \in E_{i} $, where the vector $t$ is zero
(respectively, $\, t=\frac{2}{n+2}\nabla \cdot{T} $). \item (ii)
$T(x, \{y,z\}) + T(z, \{x,y\}) + T(y, \{z,x\}) = 0$, for $x$, $y$,
$z$, eigenvectors with different eigenvalue.
\end{description}
\end{proposition}

Let $K$ be a Killing tensor of type I and let $\{e_a\}$ and
$\{\rho_a\}$ be an orthonormal basis of eigenvectors and the
corresponding eigenvalues. A straightforward calculation allows us
to write the two conditions in proposition \ref{caracter1} in terms
of $\{e_a\}$ and  $\{\rho_a\}$ obtaining, in this way:
\begin{eqnarray}
\rho_a s_{bca} + \rho_b s_{cab} + \rho_c s_{abc} & = & 0 \, ,
\qquad \qquad a,b,c \not=  \label{ike1} \\
e_a^2 e_b(\rho_a) - (\rho_b - \rho_a) s_{aab} & = & 0 \, ,
\qquad \qquad a \not= b  \label{ike2} \\
e_b(\rho_b) & = & 0         \label{ike3}
\end{eqnarray}
where $s_{abc}$ are the symmetrized rotation coefficients, $s_{abc}
= g(e_c, \{e_a, e_b\})$. If we put equations (\ref{ike1}-\ref{ike3})
in terms of the rotation coefficients we easily recover the {\em
intrinsic Killing tensor equations} obtained by Eisenhart.\cite{eis}
In order to study the metrics which admit a second order Killing
tensor, Eisenhart\cite{eis} started from these intrinsic equations
and he looked for a set of equivalent conditions involving the
eigenvectors exclusively. He considered the case when all the
eigenvalues are equal and the case with different eigenvalues and
normal principal congruences.\cite{eis} In this work we solve this
Eisenhart problem for both the Killing and conformal tensors, when
the second order tensor admits two complementary eigenspaces. We
could also start from equations (\ref{ike1}-\ref{ike3}) and similar
conditions for the conformal case, but we will choose an alternative
approach that makes the geometric properties of the eigenspaces of
the Killing and conformal tensors more evident.

Let $\rho_i$ and  $h_i$ be the eigenvalue and the projector
associated with the eigenspace $E_i$, and let $p_i$ be its
dimension. Then:
\begin{equation}  \label{tipusIa}
T = \sum \rho_i \ h_i \ ; \qquad g = \sum h_i \ ;  \qquad  \tr h_i = p_i
\end{equation}
With this notation, the second statement of lemma \ref{lem-regular}
becomes, $\, h_i({\rm d}\rho_i - t) = 0 $ and, consequently,
\begin{equation}  \label{t}
t = \sum h_i({\rm d}\rho_i)
\end{equation}
On the other hand, by projecting condition (i) in proposition
\ref{caracter1} on every eigenspace $E_j$ one obtains:
\begin{equation}
(\rho_j - \rho_i) h_j(\{x,y\}) = g(x,y) h_j({\rm d}\rho_i - t)
\end{equation}
So, if $v_i$ denotes the projection on the orthogonal space
$E_i^{\perp}$, one has:
\begin{equation}
v_i(\{x,y\}) = g(x,y) \sum_{j \not= i} \frac{1}{\rho_j -
\rho_i} h_j({\rm d}\rho_i - t)
\end{equation}
for every $x,y \in E_i$. Then, according to lemma
\ref{lemma-umbilical} and taking into account that $t$ is zero for a
Killing tensor and it can be written as (\ref{t}) for a conformal
one, we arrive to the following:

\begin{theorem}  \label{theo1}
Let $T$ be a symmetric 2-tensor of type I and let $h_{i}$ be the
projector corresponding to the eigenvalue $\rho_i$. Then, $T$ is a
Killing or a conformal tensor if, and only if,
\begin{description}
\item (i) The eigenspaces are umbilical subspaces, that is, their
second fundamental form can be written as: $S_i = \frac{1}{2} h_i
\otimes \textit{\textbf a}_i$. \item (ii) For every eigenspace the
trace of its second fundamental form $\displaystyle \tr S_i =
\frac{p_i}{2} \textit{\textbf a}_i$ satisfies
\begin{eqnarray}
\hspace{-1cm} \textit{\textbf a}_i = \sum_{j \not= i}
\frac{1}{\rho_j - \rho_i} h_j({\rm d} \rho_i) \, ,  \qquad \qquad
h_i({\rm d}\rho_i) = 0 \, ,
  \qquad \quad  \mbox{for a Killing tensor}  \label{trKT} \\
\hspace{-1cm} \textit{\textbf a}_i = - \sum_{j \not= i} h_j({\rm
d}\ln|\rho_i - \rho_j|) \, , \qquad  \ \sum p_i \rho_i = 0 \, ,
\qquad  \mbox{for a conformal tensor}  \label{trKT}
\end{eqnarray}
\item (iii) $T(x, \{y,z\}) + T(z, \{x,y\}) + T(y, \{z,x\}) = 0$,
for $x$, $y$, $z$, eigenvectors with different eigenvalues.
\end{description}
\end{theorem}

The first condition of this theorem gives a geometric property
involving the eigenvectors exclusively: every eigenspace is an
umbilical subspace. Thus, it offers a decoupled equation that
partially solves the Eisenhart problem. In next section we will
analyze the other two conditions in theorem \ref{theo1} for the case
of two complementary eigenspaces. The last condition makes no sense
in this case and we will see that the second one can be easily
decoupled.

\section{\small{GEOMETRY OF KILLING AND CONFORMAL TENSORS
OF TYPE}  \large{$\; p+q$}}  \label{sec-p+q}

A particular case of type I second order tensors are those having
two complementary eigenspaces of dimensions $p$ and $q=n-p$. So, a
$p+q$ almost-product structure $(V,H)$ is associated with these
tensors, and we say that they are of type $p+q$. If $v$ and $h$ are
the projectors onto the eigenspaces and $\alpha$ and $\beta$ are the
eigenvalues, such a tensor takes the form $T=\alpha \ v + \beta \
h$. In this case the previous theorem can be stated concisely in
terms of the canonical elements $(v,h; \alpha, \beta)$ as:

\begin{proposition} \label{k1}
A symmetric 2-tensor $K= \alpha v + \beta h$ of type $p+q$ is a
Killing tensor if, and only if, the following conditions hold:

(i) The eigenstructure $(V,H)$ is umbilical, that is, the second
fundamental forms can be written as:
\begin{equation} \label{k-u}
S_{v} = \frac12 v \otimes \textit{\textbf a} \, , \qquad \qquad
S_{h} = \frac12 h \otimes \textit{\textbf b}
\end{equation}

(ii) The traces of the second fundamental forms, $\displaystyle \tr
S_{v}= \frac{p}{2}\textit{\textbf a}$ and  $\displaystyle \tr S_{h}=
\frac{q}{2}\textit{\textbf b}$,  and the eigenvalues $\alpha$,
$\beta$ are related by
\begin{equation} \label{k0}
\textit{\textbf a} = \frac{1}{\beta- \alpha} {\rm d}\alpha \, ,
\qquad  \qquad  \textit{\textbf b} = \frac{1}{\alpha - \beta} {\rm
d}\beta
\end{equation}
\end{proposition}

\noindent A similar result takes place for conformal tensors as the
following proposition says.

\begin{proposition} \label{ck1}
A traceless symmetric 2-tensor $P=\alpha (q v - p h)$ of type $p+q$
is a conformal tensor if, and only if, the following conditions
hold:

(i) The eigenstructure $(V,H)$ is umbilical, that is, the second
fundamental forms can be written as:
\begin{equation} \label{ck-u}
S_{v} = \frac12 v \otimes \textit{\textbf a} \, , \qquad \qquad
S_{h} = \frac12 h \otimes \textit{\textbf b}
\end{equation}

(ii) The traces of the second fundamental forms, $\displaystyle \tr
S_{v}= \frac{p}{2}\textit{\textbf a}$ and  $\displaystyle \tr S_{h}=
\frac{q}{2}\textit{\textbf b}$, and the scalar $\alpha$ are related
by
\begin{equation} \label{ck2}
\textit{\textbf a} +  \textit{\textbf b} = - {\rm d} \ln |\alpha|
\end{equation}
\end{proposition}

It is worth remembering that, for the space-time 2+2 case, the
umbilical nature of the structure is equivalent to the geodesic and
shear-free character of its two null principal directions.\cite{fsD}
Consequently, the above propositions generalize some results for the
space-time Killing and conformal tensors of type 2+2 (see Ref. 5,
theorem 35.4) to an arbitrary dimension $n$ and an arbitrary type
$p+q$. Now we want to remark that the covariant formalism used here
allow us to accomplish the second step in the Eisenhart method: the
characterization of the Killing and conformal tensors in terms of
their eigenspaces.

The characterization  of a $p+q$ Killing or conformal tensor
presented in the propositions above involves the structure tensor
(conditions (i) and (ii)) and the eigenvalues (condition (ii)). The
next step consists of removing the eigenvalues in order to obtain
the conditions that an almost product structure must satisfy in
order to be the eigenstructure of a Killing or a conformal tensor.
Condition (ii) of proposition \ref{k1} can be written as
\begin{equation} \label{aux1}
(\alpha - \beta ) \textit{\textbf a} = - {\rm d} \alpha ; \qquad
(\alpha - \beta ) \textit{\textbf b} =  {\rm d} \beta
\end{equation}
Then we have $(\beta - \alpha) \ (\textit{\textbf a}+\textit{\textbf
b}) =  {\rm d} (\alpha - \beta )$. If we differentiate (\ref{aux1})
and make the substitution of ${\rm d} (\alpha - \beta )$ we get
\begin{equation} \label{aux2}
{\rm d} \textit{\textbf a} + \textit{\textbf a} \wedge
\textit{\textbf b} =0 \, , \qquad {\rm d} \textit{\textbf b} +
\textit{\textbf b} \wedge \textit{\textbf a} = 0
\end{equation}
Conversely, if $\textit{\textbf a}$, $\textit{\textbf b}$ satisfy
equations (\ref{aux2}), two functions $x$, $y$ exist such that
$$\textit{\textbf a}+\textit{\textbf b} = {\rm d} x ,
\qquad \textit{\textbf a}-\textit{\textbf b} = e^{x} {\rm d} y $$
Then, taking $\alpha = e^{-x} - y$ and $\beta = -e^{-x} - y$,
equation (\ref{aux1}) is satisfied and $\, K= \alpha \, v + \beta \,
h \,$ is a Killing tensor provided that (\ref{k-u}) holds. The
freedom in choosing $x$ and $y$ leads to the family of Killing
tensors $\, C \, K + D \, g\,$, $C$ and $D$ being arbitrary
constants.

In the same way, condition (\ref{ck2}) for a conformal tensor
implies that $ {\rm d} (\textit{\textbf a}+\textit{\textbf b})=0$.
Conversely, if $ {\rm d} (\textit{\textbf a}+\textit{\textbf b})=0$,
a function $x$ exists such that $\textit{\textbf a}+\textit{\textbf
b} = {\rm d} x$. Then, the traceless tensor $\, P= e^{-x} \, (q \, v
- p \, h )\,$ is a conformal Killing tensor provided that
(\ref{ck-u}) holds. The freedom in choosing $x$ leads to the family
$\, C P\, $, $C$ being an arbitrary constant.
Thus, we have obtained:

\begin{theorem} \label{teorema2}
The necessary and sufficient conditions for a $p+q$ almost-product
structure $(V,H)$ to be the eigenstructure of a Killing or a
conformal tensor are:

(i) $(V,H)$ is umbilical, that is, the second fundamental forms
take the expression:
\begin{equation}  \label{k-c-um}
S_{v} = \frac12 v \otimes \textit{\textbf a} \, , \qquad \qquad
S_{h} = \frac12 h \otimes \textit{\textbf b}
\end{equation}

(ii) The traces, $\displaystyle \tr S_{v}=
\frac{p}{2}\textit{\textbf a}$ and $\displaystyle \tr S_{h}=
\frac{q}{2}\textit{\textbf b}$, of the second fundamental forms
satisfy:
\begin{equation} \label{trazak}
{\rm d} \textit{\textbf a} + \textit{\textbf a} \wedge
\textit{\textbf b} =0 \, , \qquad {\rm d} \textit{\textbf b} +
\textit{\textbf b} \wedge \textit{\textbf a} = 0 \quad \qquad
\mbox{for Killing tensors}
\end{equation}
\begin{equation} \label{trazack}
\qquad \qquad {\rm  d } (\textit{\textbf a}+\textit{\textbf b}) =0
\qquad \qquad \qquad \qquad \mbox{for conformal tensors}
\end{equation}

\noindent If {\rm (\ref{k-c-um})} and {\rm (\ref{trazak})} hold, two
functions $x$, $y$ exist such that $\textit{\textbf
a}+\textit{\textbf b} = {\rm d} x$, $\textit{\textbf
a}-\textit{\textbf b} = e^{x} {\rm d} y$. Then taking $\alpha=e^{-x}
- y$, $\beta= - e^{-x} - y$, $\, K= C(\alpha \, v + \beta \, h) + D
g $ is
a Killing tensor, $C$ and $D$ being two arbitrary constants. \\[1mm]
If {\rm (\ref{k-c-um})} and {\rm (\ref{trazack})} hold, a function
$x$ exists such that ${\rm d} x = \textit{\textbf a}+\textit{\textbf
b}$. Then, $P= C \, e^{-x} \, (q \, v - p \, h)$ is a conformal
Killing tensor, $C$ being an arbitrary constant.
\end{theorem}

This theorem offers the second step in solving the Eisenhart problem
for Killing or conformal tensors with two complementary eigenspaces.
In fact, once the eigenvalues have been removed, we have obtained
necessary and sufficient conditions involving the sole eigenspaces.
In section \ref{sec-2+2} we will see that, for the space-time 2+2
case, these conditions can be written as tensorial conditions on the
structure tensor (or on the canonical 2--form associated with the
structure). This fact allows us to give an intrinsic and explicit
characterization of the four dimensional Petrov-Bel type D
space-times admitting a Killing or a conformal tensor in section
\ref{sec-2+2-D}.

\section{\small{METRICS ADMITTING A KILLING OR A CONFORMAL
TENSOR OF TYPE} \large{$\; p+q$}} \label{sec-metrics-p+q}

In this section we show that a metric admitting a Killing or a
conformal tensor of type $p+q$ admits a canonical expression in
terms of a particular conformal metric and a specific conformal
factor. Firstly we state a corollary which trivially follows on from
propositions \ref{k1} and \ref{ck1}:

\begin{corollary} \label{cor-constants}
Let $(V,H)$ be a $p+q$ almost-product structure for the metric
tensor $g$. The following statements are equivalent:

(i) $(V,H)$ is a p+q totally geodesic almost-product structure.

(ii) $C\, v + D\, h$ is a Killing tensor, $C$ and $D$ being
arbitrary constants.

(iii) $C\, (q v - p h)$ is a conformal tensor, $C$ being an
arbitrary constant.
\end{corollary}

This corollary states that the Riemannian spaces admitting a second
order Killing tensor with constant eigenvalues are those admitting a
p+q totally geodesic structure $(V,H)$. We will show now that these
Riemannian spaces generate all the spaces admitting Killing or
conformal tensors by using an adequate conformal transformation.

The umbilical property is known to be a conformal
invariant.\cite{olga,fsD} Moreover, if we take into account the
change of the second fundamental form through a conformal
transformation,\cite{fsD} condition (\ref{ck2}) for a conformal
tensor states that the eigenstructure $(V,H)$ is minimal for the
conformal metric $\tilde{g} = |\alpha|^{-1} g$. Consequently, the
family of metrics that admit a $p+q$ conformal tensor are those that
are conformal to a metric which admits a totally geodesic $p+q$
structure. More precisely, we have:

\begin{proposition} \label{pro-can-ct}
The metrics $g$ that admit a $p+q$ conformal tensor are those that
may be written as $g = |\alpha| \tilde{g}$, where $\tilde{g}$ is a
metric admitting a totally geodesic $p+q$ structure $(V,H)$.\\[1mm]
Then the conformal tensor for $g$ is $P=C \alpha (q\,v-p\,h)$, $C$
being an arbitrary constant.
\end{proposition}

This proposition and corollary \ref{cor-constants} generalize to an
arbitrary dimension $n$ and an arbitrary type $p+q$ a result by
Hauser and Malhiot\cite{h-m-2} concerning the $2+2$ space-time case.
Moreover we also recover another known result
easily:\cite{r-e-barnes} a (contravariant) conformal tensor for a
metric is a conformal tensor for every conformally related metric.

A similar result holds for Killing tensors. In fact, the sum of
expressions (\ref{trazak}) says that ${\rm d} (a+b)=0$, which is
exactly the condition necessary for $(V,H)$ to be the eigenstructure
of a conformal tensor, and so the metric is conformal to a metric
admitting a $p+p$ totally geodesic structure. But now, the conformal
factor is not arbitrary because it must satisfy the two equations in
(\ref{trazak}). A detailed analysis of these conditions leads to:

\begin{proposition}  \label{pro-can-kt}
The metrics $g$ that admit a p+q Killing tensor are those that may
be written as $g = |\alpha - \beta| \tilde{g}$, where $\tilde{g}$ is
a metric admitting  a totally geodesic $p+q$ structure $(V,H)$, and
$\alpha$ and $\beta$ are functions such that $v({\rm d} \alpha) =0$,
$h({\rm d} \beta) =0$.\\[1mm]
Moreover, the Killing tensor for $g$ is $K= C \, (\alpha \, v
+\beta \, h) + D g $, $C$ and $D$ being arbitrary constants.
\end{proposition}

The two propositions above imply that the study of the Riemannian
spaces admitting a Killing or a conformal tensor reduces to the
study of the metrics $\tilde{g}$ admitting a totally geodesic $p+q$
structure. As proposition \ref{pro-can-ct} states, for every metric
$\tilde{g}$ of this type we obtain a metric $g$ admitting a
conformal tensor by using an arbitrary conformal factor, $g =
\Omega^2 \tilde{g}$.

Nevertheless, proposition \ref{pro-can-kt} states that the richness
of metrics admitting a Killing tensor conformally related to a
$\tilde{g}$ of this type depends on the quantity of normal
directions aligned with one of the planes of the structure. This
fact induces a classification of the metrics  admitting a totally
geodesic $p+q$ structure.

In the more regular metrics no aligned normal direction exists and
only constant conformal factors can be considered, the Killing
tensor then have constant eigenvalues.

The more degenerate class corresponds to the product metrics
$\tilde{g} = \tilde{v} + \tilde{h}$, $\tilde{v}_{AB}(x^C)$ and
$\tilde{h}_{ij}(x^k)$ being two arbitrary $p$ and $q$ dimensional
metrics, respectively; then, the available conformal factors are
$\Omega^2 = |\alpha - \beta|$, $\alpha(x^k)$ and $\beta(x^C)$ being
arbitrary functions depending on the product coordinates and they
coincide with the Killing tensor eigenvalues.

An intermediate situation occurs when, for example, only one normal
aligned direction exists on each plane. Then, through the adequate
conformal transformation we can obtain a metric admitting a Killing
tensor with non-constant eigenvalues. In dealing with $2+2$
space-time Killing tensors this case leads to the Hauser and
Malhiot\cite{h-m-1,h-m-2} canonical form for the metric.

\section{\small{KILLING AND CONFORMAL TENSORS OF TYPE}
{\large $\; 1+(n-1)$}} \label{sec-1+q}

Let us consider the case of a $1+(n-1)$ structure $(V,H)$ defined
by the unitary direction $u$ ($u^2 = \epsilon = \pm 1$) and its
orthogonal complement. Then $g = v + h$ where $v=\epsilon \ u
\otimes u $ and $h = g - \epsilon \ u \otimes u $.  In terms of
the usual kinematic coefficients of $u$ ($\nabla u = \epsilon \ u
\otimes \dot{u} + \frac{1}{n-1} \theta \ h + \sigma + \Omega$) the
(generalized) second fundamental forms are
\begin{equation}
Q_v = u \otimes u \otimes \dot{u} ; \qquad
Q_h = - \epsilon \ \Big( \frac{1}{n-1} \theta \ h + \sigma + \Omega \Big)
\otimes u
\end{equation}
The condition for $(V,H)$ to be an umbilical structure just states
$\sigma =0$, and then:
\begin{equation} \label{sffumbi}
S_v =  u \otimes u \otimes \dot{u} ; \qquad
S_h = - \epsilon \ \frac{1}{n-1} \theta \ h \otimes u
\end{equation}
Thus taking into account theorem \ref{teorema2}, we find that the
necessary and sufficient condition for $u$ to define the
eigenstructure of a conformal tensor is
\begin{equation} \label{ck11}
\sigma = 0  \, , \qquad {\rm d} ( \dot{u} - \frac{\theta}{n-1} \  u) =0
\end{equation}
But these conditions state that $u$ defines the direction of a
conformal Killing vector.\cite{ehlers} Thus, we have:

\begin{proposition}
A $1+(n-1)$ structure defined by the unitary direction $u$ is the
eigenstructure of a conformal Killing tensor if, and only if, $u$
defines the direction of a conformal Killing vector, that is, it
satisfies {\rm (\ref{ck11})}.
\end{proposition}

This proposition implies that every traceless conformal tensor of
type $1+(n-1)$ is the traceless part of $\,\xi \otimes \xi$, $\xi$
being a Killing conformal vector. In other words: {\it every
$1+(n-1)$ conformal tensor is reducible}.

A similar procedure allows us to characterize the fact that $u$
defines the eigenstructure of a $1+(n-1)$ Killing tensor. But in
this case we find that it is not, necessarily, reducible. Indeed,
taking into account (\ref{sffumbi}) the condition (\ref{trazak})
of theorem \ref{teorema2} is equivalent to
$${\rm d} ( \dot{u} - \frac{\theta}{n-1} \  u) =0 ,
\qquad \theta {\rm d} u + {\rm d} \theta \wedge u + 2 \epsilon
\theta \ u \wedge \dot{u}=0 $$
When $\theta =0$ these equations hold if ${\rm d} \dot{u} =0$,
that is, if $u$ defines the direction of a Killing vector. On the
contrary, if $\theta \neq 0$, the second equation implies ${\rm d}
u \wedge u =0$, and so ${\rm d} u = \epsilon \ u \wedge \dot{u}$.
In this case, $u$ defines the direction of a normal conformal
Killing vector and the second equation can be written as
\begin{equation} \label{kt2}
\displaystyle {\rm d} (\theta^{\frac{1}{3}} u) = 0 \, .
\end{equation}
These results are summarized in the following

\begin{proposition}
The $1+(n-1)$ structure defined by the unitary direction $u$ is
the eigenstructure of a  Killing tensor if, and only if, one of
the following conditions hold:

(i) $u$  defines the direction of a Killing vector, that is, it
satisfies $\sigma =0 = \theta$, ${\rm d} \dot{u}=0$.

(ii) $u$ defines the direction of a  normal conformal Killing vector
with integrant factor $\theta^{1/3}$, that is, it satisfies
equations {\rm (\ref{ck11})} and {\rm (\ref{kt2})}.
\end{proposition}

This proposition shows that we can distinguish two classes of
Killing tensors of type $1+(n-1)$. On the one hand, we have the
reducible ones, that is, those that can be written as $\, \xi
\otimes \xi \,+\, B g$, $\xi$ being a Killing vector and $B$ an
arbitrary constant. On the other hand, a class of irreducible
Killing tensors that can be obtained from normal conformal Killing
vectors. This last class has been considered by
Koutras\cite{koutras} and Rani {\it et al.}\cite{r-e-barnes}

The results in the previous section allow us to give the canonical
form for the metric tensors admitting irreducible Killing tensors of
type $1+(n-1)$. Indeed, as the eigenstructure is integrable, the
metric will be conformally related to a $1+(n-1)$ product metric.
Moreover proposition \ref{pro-can-kt} gives the conformal factor.
Finally, we can state:

\begin{proposition}
The metrics admitting a irreducible Killing tensor of type
$1+(n-1)$ are those that may be written as
\begin{equation}
g = |\alpha(x^i) - \beta(x^0)| [\epsilon \, {\rm d}x^0 \otimes
{\rm d}x^0 + \gamma(x^i)]
\end{equation}
where $\gamma(x^i)$ is an arbitrary ($n-1$)--dimensional metric.

The Killing tensor is then given by $\, C |\alpha - \beta| [\epsilon
\alpha \, {\rm d}x^0 \otimes {\rm d}x^0 + \beta \gamma(x^i)] + D g$,
$C$ and $D$ being arbitrary constants.
\end{proposition}

\section{\small \hspace*{-3mm}{SPACE-TIME KILLING AND CONFORMAL TENSORS
OF TYPE} \large{$[(11)(11)]$}}  \label{sec-2+2}

Let $T$ be a Killing or a conformal tensor of type $[(11)(11)]$ in
an oriented four dimensional space-time $(V_4,g)$ of signature
$(-+++)$. Then $T$ has two eigenspaces: a time-like two-plane $V$
and its space-like orthogonal complement $H$. The almost-product
eigenstructure $(V,H)$ is determined by the {\it canonical} unitary
2-form $U$, volume element of the time-like plane $V$. Then, the
respective projectors are $v=U^2$ and $h = -(*U)^2$, where $U^2 = U
\times U= \tr_{23} U \otimes U$ and $*$ is the Hodge dual operator.

In order to study the geometric properties of a $2+2$ structure it
is useful to introduce the self-dual unitary 2--form ${\cal U}
\equiv \frac{1}{\sqrt{2}} (U - {\rm {i}} *U )$ associated with
$U$. The metric on the self-dual 2--forms space is ${\cal G} =
\frac{1}{2} ( G - {\rm i} \eta )$, where $\eta$ is the metric
volume element of the space-time, $G=\frac{1}{2} g \wedge g$ is
the metric on the 2--forms space, and $\wedge$ denotes the
double-forms exterior product, $(A \wedge B)_{\alpha \beta \mu
\nu} = A_{\alpha \mu} B_{\beta \nu} + A_{\beta \nu} B_{\alpha \mu}
- A_{\alpha \nu} B_{\beta \mu} - A_{\beta \mu} B_{\alpha \nu}$.
Then, we can consider some first order differential concomitants
of $U$ that determine the geometric properties of the structure.
Indeed, if $i(\cdot)$ denotes the interior product and $\delta$
the exterior codifferential, $\delta = *d*$, we have the following
lemma:\cite{fsD}

\begin{lemma}  \label{lem-imu}
Let us consider the 2+2 structure defined by ${\cal U}=\frac{1}{\sqrt{2}}
(U - i *U)$. Then: \\
(i) The traces of the second fundamental forms take the
expression:
\begin{equation}  \label{traza}
\tr Q_v = \textit{\textbf a}[U] \equiv - i (\delta *U) *U ; \qquad
\tr Q_h = \textit{\textbf b}[U] \equiv i (\delta U) U ;
\end{equation}
(ii) The structure is umbilical, if, and only if,
\begin{equation}
\Sigma[U] \equiv \nabla {\cal U} - i(\delta {\cal U}){\cal U}
\otimes {\cal U} - i(\delta {\cal U}){\cal G}=0  \label{umb}
\end{equation}
\end{lemma}

With this notation, we can write the intrinsic equations in
propositions \ref{k1} and \ref{ck1} for the case of Killing or
conformal tensors of type $[(11)(11)]$ by using the eigenvalues and
the canonical two-form $U$ exclusively:

\begin{proposition} \label{pro-2+2-ct}
The traceless symmetric tensor $P= \alpha \,[U^2 + (*U)^2]$ is a
conformal tensor if, and only if, the canonical elements $\{\alpha,
U\}$ satisfy {\rm (\ref{umb})} and:
\begin{equation}
-{\rm d} \ln|\alpha| = \Phi[U] \equiv   i(\delta U) U - i(\delta
*U) *U. \label{pre-max-alpha}
\end{equation}
\end{proposition}

\begin{proposition} \label{pro-2+2-kt}
The symmetric tensor $K= \alpha \,U^2 +  \beta \, (*U)^2$ is a
Killing tensor if, and only if, the canonical elements $\{\alpha,
\beta, U\}$ satisfy {\rm (\ref{umb})}, {\rm (\ref{pre-max-alpha})}
and:
\begin{equation}
{\rm d} \alpha = (\alpha - \beta)  i(\delta *U) *U.
\end{equation}
\end{proposition}

This last proposition is the tensorial version of the intrinsic
equations for a Killing tensor that are known in Newmann-Penrose
formalism (Ref. 5, theorem 35.4). Now we can easily write the
conditions in theorem \ref{teorema2} in terms of the canonical
two-form $U$, that is, we obtain the characterization of the Killing
and conformal tensor in the sole variable $U$:

\begin{theorem} \label{teorema3}
The $2+2$ structure defined by the unitary simple 2-form $U$ is
the eigenstructure of a conformal tensor if, and only if, $U$
satisfies:
\begin{eqnarray}
\Sigma[U] \equiv \nabla {\cal U} - i (\delta {\cal U})
{\cal U} \otimes {\cal U} - i(\delta {\cal U}) {\cal G} =0, \label{umbilical-1}\\
{\rm d} \Phi[U] \equiv {\rm d} \Big[ i(\delta U) U - i(\delta *U)
*U \Big]=0. \label{pre-max-1}
\end{eqnarray}
If these conditions hold, a function $\alpha$ exists such that
$\Phi[U] = -{\rm d} \ln|\alpha|$. Then, the conformal tensor is $P=
C \, \alpha \,[U^2 + (*U)^2]$, $C$ being an arbitrary constant.
\end{theorem}

\begin{theorem} \label{teorema4}
The $2+2$ structure defined by the unitary simple 2-form $U$ is
the eigenstructure of a Killing tensor if, and only if, $U$
satisfies:
\begin{eqnarray}
\Sigma [U] \equiv \nabla {\cal U} - i (\delta {\cal U})
{\cal U} \otimes {\cal U} - i(\delta {\cal U}) {\cal G} =0,  \label{umbilical-2}\\
{\rm d} \Phi[U] \equiv {\rm d} \Big[ i(\delta U) U - i(\delta *U)
*U \Big]=0,  \label{pre-max-2} \\
{\rm d}  i(\delta U) U = i(\delta U) U \wedge  i(\delta *U) *U.
\label{Killing-condition}
\end{eqnarray}
If these conditions hold, two functions $\alpha$ and $\beta$ exist
such that $\Phi[U] = -{\rm d} \ln|\alpha-\beta|$ and $2 (\alpha-
\beta)[i(\delta U) U + i(\delta *U) *U] = {\rm d} (\alpha + \beta)$.
Then, the Killing tensor is $K = C [\alpha \, U^2  -  \beta \,
(*U)^2] + D g $, $C$ and $D$ being two arbitrary constants.
\end{theorem}

It is worth pointing out that the first order differential
properties of a $2+2$ structure admit a kinematical
interpretation\cite{cf} and, in particular, the umbilical condition
(\ref{umbilical-1},\ref{umbilical-2}) equivalently implies that the
two principal null directions of the structure are geodesic and
shear-free congruences.\cite{fsD} Thus, we recover a known result
obtained independently by Hauser and Malhiot\cite{h-m-1} and by
Collinson.\cite{collinson}

On the other hand, condition (\ref{pre-max-1}) states that the
structure is pre-Maxwellian.\cite{debever,fsY} Then, taking into
account the study of these structures given in,\cite{fsY} we have:

\begin{corollary}
The $2+2$ traceless tensor $P = \alpha (v-h)$ is a conformal
tensor if, and only if, $T= \alpha^{-2} (v-h)$ is a conservative
Maxwell-Minkowski energy tensor and the principal directions of
the associated electromagnetic field are geodesic and shear-free
congruences.
\end{corollary}

\section{\small  PETROV-BEL TYPE D SPACE-TIMES ADMITTING KILLING
OR CONFORMAL TENSORS} \label{sec-2+2-D}

The results in previous sections help us to characterize
intrinsically and explicitly some families of metrics. More
precisely, in this section: (i) we obtain necessary and sufficient
conditions on the metric concomitants for a four dimensional
space-time to be a Petrov-Bel type D solution admitting a $2+2$
Killing or conformal tensor and, when they hold, (ii) we give an
algorithm to determine these tensors.

In the previous section we have characterized the $2+2$ Killing and
conformal tensors in terms of the volume element $U$ of their
time-like eigen-plane. Moreover, for the case of Petrov-Bel type D
metrics, this 2-plane determines the Weyl principal structure and,
consequently, $U$ can be obtained from the Weyl tensor. The
intrinsic and explicit characterization of type D solutions and the
covariant obtaining of the Weyl canonical bivector have been given
in Ref. 9. Consequently we can state the following invariant
characterizations:

\begin{proposition} \label{prop-D-umbilical}
A Petrov-Bel type D metric admits a conformal tensor if, and only
if, the Weyl principal null directions define geodesic shear-free
congruences and the Weyl canonical 2-form satisfies {\rm
(\ref{pre-max-2})}.

A Petrov-Bel type D metric admits a Killing tensor if, and only if,
the Weyl principal null directions define geodesic shear-free
congruences and the Weyl canonical 2-form satisfies {\rm
(\ref{pre-max-2})} and {\rm (\ref{Killing-condition})}.
\end{proposition}

Finally, taking into account the algebraic results for Petrov-Bel
type D metrics quoted above (see Ref. 9), we obtain from theorems
\ref{teorema3} and \ref{teorema4} the explicit expression of the
conditions in proposition \ref{prop-D-umbilical} and the algorithm
for obtaining the conformal or Killing tensors:

\begin{theorem} \label{theorem-st1}
Let ${\cal W} \equiv {\cal W}(g) = \frac12 (W(g) - \ci *W(g))$ and
${\cal G} \equiv {\cal G}(g) = \frac12 (\frac12 g \wedge g - \ci
\eta(g))$ the self-dual Weyl tensor and self-dual metric associated
with a space-time metric $g$, and let us take the metric
concomitants:
\begin{eqnarray}
\rho \equiv  - \frac{{\rm \tr}{\cal W}^3}{{\rm \tr} {\cal W}^2} \, ,
\qquad \qquad {\cal S} \equiv \frac{1}{3 \rho} \left({\cal W} - \rho
\, {\cal G}\right) \, , \qquad  \quad {\cal U} \equiv \frac{{\cal
S}({\cal
X})}{\sqrt{{\cal S}({\cal X},{\cal X})}} \, ,  \label{concomitants-1}\\[2mm]
\Sigma  \equiv \nabla {\cal U} - i (\delta {\cal U}) {\cal U}
\otimes {\cal U} - i(\delta {\cal U}) {\cal G} \, , \qquad \qquad
\qquad \qquad \qquad \qquad \qquad
\\[2mm]
U \equiv \sqrt{2}Re\{{\cal U}\} \, , \quad  \qquad  \textit{\textbf
a} \equiv - i (\delta *U) *U \, , \qquad \textit{\textbf b}  \equiv
i (\delta U) U \, . \qquad \quad
\end{eqnarray}
where ${\cal X}$ is an arbitrary self-dual bivector.

The necessary and sufficient conditions for $g$  to be a Petrov-Bel
type D solution admitting a $2+2$ conformal tensor are:
\begin{equation}
\rho  \not= 0 \, , \qquad {\cal S}^2 + {\cal S} = 0 \, , \qquad
\Sigma = 0 \, , \qquad {\rm d}(\textit{\textbf a} + \textit{\textbf
b})=0 \label{conforme-killing}
\end{equation}
When {\rm (\ref{conforme-killing})} hold, a function $\alpha$ exists
such that $-{\rm d} \ln|\alpha| = \textit{\textbf a} +
\textit{\textbf b}$. Then, the conformal tensor is $P= C \, \alpha
\,[U^2 + (*U)^2]$, $C$ being an arbitrary constant.

The necessary and sufficient conditions for $g$  to be a type D
solution admitting a $2+2$ Killing tensor are {\rm
(\ref{conforme-killing})} and:
\begin{equation}
{\rm d}\textit{\textbf b} + \textit{\textbf b} \wedge
\textit{\textbf a} = 0              \label{killing-tensor}
\end{equation}
When {\rm (\ref{conforme-killing})} and {\rm (\ref{killing-tensor})}
hold, two functions $\alpha$ and $\beta$ exist such that $-{\rm d}
\ln|\alpha-\beta| = \textit{\textbf a} + \textit{\textbf b}$ and $
{\rm d} (\alpha + \beta) = 2 (\alpha- \beta)[\textit{\textbf b} -
\textit{\textbf a}]$. Then, the Killing tensor is $K = C [\alpha \,
U^2 - \beta \, (*U)^2] + D g $, $C$ and $D$ being two arbitrary
constants.

\end{theorem}

For Petrov-Bel type D solutions with a vanishing Cotton tensor (the
Weyl tensor is divergence-free) the Bianchi identities take the
expression:\cite{fsD}
\begin{equation} \label{divcero}
\nabla {\cal U} = i(\delta {\cal U}) [{\cal U} \otimes {\cal U} +
{\cal G} ] \ ; \ \quad i (\delta{\cal U} ) {\cal U} = \frac{1}{3} \
\mbox{\rm d} \ln{\rho}
\end{equation}
where ${\cal U}$ is the Weyl canonical bivector and $\rho$ the
double Weyl eigenvalue. The real part of the second equation in
(\ref{divcero}) states:
\begin{equation}
\frac23 \, {\rm d} \ln|\rho| = \Phi[U] \equiv   i(\delta U) U -
i(\delta *U) *U. \label{pre-max-rho}
\end{equation}
Thus, the principal structure of a type D divergence-free Weyl
tensor is umbilical and pre-Maxwellian and, as a consequence of
theorem \ref{teorema3}, it is the structure of a conformal tensor.
Moreover, in this case the eigenvalue of the conformal tensor can be
obtained algebraically from the Weyl eigenvalues if we take into
account proposition \ref{pro-2+2-ct}. Thus, we have:
\begin{theorem} \label{theorem-st2}
Every Petrov-Bel type D solution with vanishing Cotton tensor admits
a conformal tensor. Let $\rho$, ${\cal S}$ and ${\cal U}$ be the
Weyl concomitants given in {\rm (\ref{concomitants-1})}. Then:

(i) These space-times are characterized by the conditions:
\begin{equation}
\rho  \not= 0 \, , \qquad {\cal S}^2 + {\cal S} = 0 \, , \qquad
\delta W = 0 \label{D-Cotton}
\end{equation}

(ii) The conformal tensor is given by
\begin{equation}
P=  C \, |\rho|^{-2/3} \, {\cal U} \times \tilde{{\cal U}}
\end{equation}
\end{theorem}
This theorem generalizes the result about the existence of conformal
tensors in Petrov-Bel type D vacuum solutions (see Ref. 5, theorem
35.2).

We finish with two comments. The characterization of the Killing or
conformal tensors in terms of their underlying structure has allowed
us to give an explicit and intrinsic labeling of the Petrov-Bel type
D space-times admitting Killing or conformal tensors, as well as to
generalize some known results on the existence of these symmetries.
Furthermore, our Eisenhart-like approach to the Killing and
conformal tensor may also be useful in analyzing and extending other
properties. For example, it is known that all type D vacuum
solutions that admit a Killing tensor, also admit a Killing-Yano
tensor.\cite{collinson,ste} Our result here and those given in Ref.
22 allow us to generalize this property. This question and other
related topics will be considered elsewhere.\cite{fs-Dbuit}

Our study of the geometry of the Killing and conformal tensors and
the canonical expressions of the metric tensor in terms of this
geometry can be applied, in particular, to n-dimensional Lorentzian
metrics. We know that, for four dimensional Petrov-Bel type D
space-times, this underlying geometry is closely related with the
Weyl tensor and, this fact allows us to determine the $2+2$ Killing
and conformal tensors (see theorems 5 and 6). The generalization of
these results to higher dimensions is an open problem that could be
fruitful in some classes of the Weyl tensor. But this study will
require a further analysis of the Weyl classification in higher
dimensions.\cite{smet,coley}

\section*{\small ACKNOWLEDGEMENTS}
This work has been partially supported by the Spanish Ministerio
de Educaci\'on y Ciencia, MEC-FEDER project AYA2003-08739-C02-02.

\end{document}